\documentclass[pre,preprint,onecolumn,a4paper,showkeys,showpacs]{revtex4}

\usepackage{graphicx}

\begin{document}

\title{A one-dimensional model for the growth of CdTe quantum dots on Si substrates}

\author{S. C. Ferreira Jr.}\email{silviojr@ufv.br.}\thanks{~~~\textit{Present Address}: Departamento de F\'{\i}sica, Universidade Federal de Pernambuco, 50670-901, Recife, PE, Brazil.\\}

\author{S. O. Ferreira}
\affiliation{Departamento de F\'{\i}sica, Universidade Federal Vi\c{c}osa, 36571-000, Vi\c{c}osa, MG, Brazil}


\begin{abstract}
Recent experiments involving CdTe films grown on Si(111) substrates by hot wall epitaxy revealed features not previously observed [S. O. Ferreira \textit{et al.}, J. Appl. Phys. \textbf{93}, 1195 (2003)]. This system, which follows the Volmer-Weber growth mode with nucleation of isolated 3D islands for less than one monolayer of evaporated material, was described by a peculiar behavior of the quantum dot (QD) size distributions. In this work, we proposed a kinetic deposition model to reproduce these new features. The model, which includes thermally activated diffusion and evaporation of CdTe, qualitatively reproduced the experimental QD size distributions. Moreover, the model predicts a transition from Stranski-Krastanow growth mode at lower temperatures to Volmer-Weber growth mode at higher ones characterized through the QD width distributions.
\end{abstract}


\maketitle

\section{Introduction}

The growth of surfaces by distinct deposition techniques as well as the theoretical understanding
of these processes are among the most challenging topics in Physics [1,2]. In
special, semiconductor quantum dot structures have attracted a lot of attention in recent
years due to their exciting electronic properties and potential applications in optoelectronic
devices. Almost the totality of the semiconductor nanostructures have been grown by using
the transition from two- to three-dimensional growth regime driven by the strain energy accumulated in the epitaxial layer [3-9], the Stranski-Krastanow (SK) growth mode. Ferreira
\textit{et al}. have shown that the growth of CdTe quantum dots (QDs) on Si(111) substrates using hot wall epitaxy (HWE) follows the Volmer-Weber (VW) growth mode. This system exhibits 
nucleation of isolated 3D CdTe islands even for just 0.6 monolayers of deposited material
[10]. According to this work, the central difference between SK and VW growth modes
lies on the behavior of the dot size and density distribution as functions of temperature.
The authors have investigated CdTe samples grown on Si(111) for distinct growth times
and substrate temperatures. Figures 1(a)-(c) show samples grown with different times, namely 1.6, 3.2 and 6.4 monolayers (ML) of evaporated material, at a temperature
200 $^\circ$C. Also, the effects of temperature are illustrated in figures 1(d)-(f), in which samples with 1.6 ML grown at 200, 250 and 300 $^\circ$C are shown. In figure 2, one can observe a very peculiar behavior of the density and size distribution of the quantum dots when the temperature increases. This feature leads to a completely different growth dynamics of this system under VW growth mode. This difference probably is due
to the absence of the wetting layer in this growth mode.

\begin{figure*}[hbt]
\begin{center}
\includegraphics[clip=true,width=10.0cm]{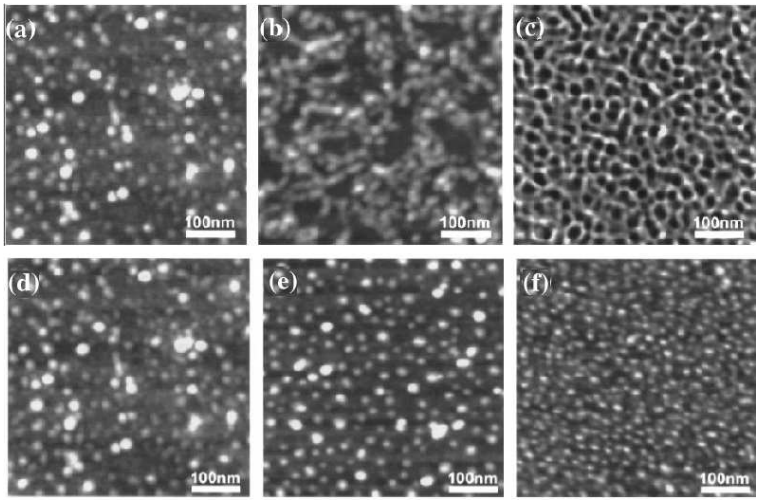}
\end{center}
\vspace{-0.5cm}
\caption{AFM images of three CdTe samples grown at 200$~^\circ$C with distinct amounts of evaporated CdTe (a) 1.6 ML, (b) 3.2 ML, and (c) 6.4 ML (top), and three samples with 1.6ML grown at different substrate temperatures (d) $200~^\circ$C, (e) $250~^\circ$C, (f) $300~^\circ$C (bottom). (This results were previously reported in reference [10].)}
\label{japfig2}
\end{figure*}

\begin{figure}[hbt]
\begin{center}
\includegraphics[clip=true,width=9cm]{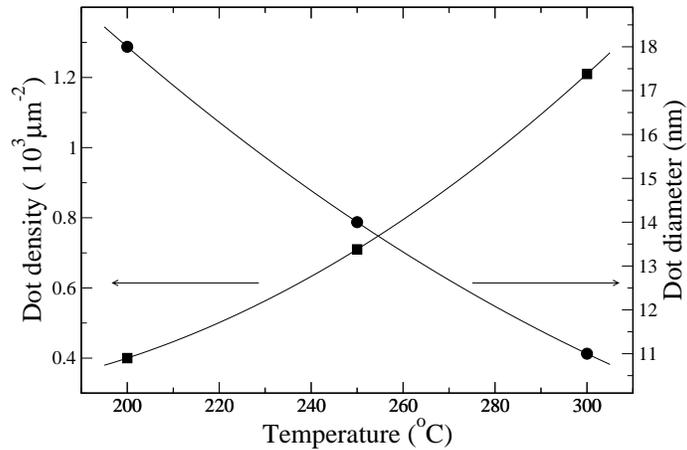}
\end{center}
\vspace{-0.5cm}
\caption{Dot density (squares) and diameter (circles) as functions of temperature. These data were taken from reference \cite{Suka}.}
\label{japfig6}
\end{figure}

Intense effort has been devoted to propose models which describe the interface evolution
of different deposition processes. In general, two approaches are used in these models: the
first one is based on continuous equations [11-14] and the second one makes use of a discrete
formulation in which the atomistic features are taken into account. In the present paper we
used the second one. Several models for epitaxial growth and their respective scaling properties were exhaustively investigated [13-19]. In particular, models focusing the growth of three-dimensional islands in heteroepitaxial films were considered [20-23]. Jensen \textit{et al.} [20] studied a model including deposition, diffusion, evaporation, aggregation, and surface defects. In this work, they evaluated the island size distributions for distinct evaporation rates (associated to the temperature) but, the qualitative behaviors obtained for 2D islands [20] do not agree with those observed in CdTe/Si experiments [10]. In reference [23], a model for CdTe films grown on Si(001) substrates through pulsed laser deposition (PLD) was considered.
In these simulations, the VW growth mode was observed but, self-assembled QDs were
not reported. The central difference between PLD and HWE methods and, consequently,
between the model of reference [23] and the one presented in this paper, is that the atoms
are deposited in the dissociated form (isolated Cd e Te atoms) in the former, whereas entire
molecules are deposited in the later.

In this paper we present a model for the growth of CdTe films on Si(111) substrates including CdTe diffusion and evaporation. The main features of the model are the distinct rules for diffusion and evaporation  of CdTe on the initial Si substrate and of CdTe on previously deposited CdTe layers. The paper outline is the following. Section II describes the model and
the procedures used in the simulations. In section III, the simulation results are presented
and discussed in the light of the experiments. Some conclusions are drawn in section IV.

\section{Model and methods}
\label{model}
The Si(111) substrate is represented by one-dimensional lattices with $L$ sites and periodic
boundary conditions. The Si adatoms do not diffuse or evaporate, i. e., they are
inert. Three processes are included in the CdTe dynamics: \textit{deposition}, \textit{diffusion} and \textit{evaporation}. The substrate height at the site $j$, i.e., the number of CdTe adatoms on this site, is
represented by $h_j$. In following paragraphs, each process is described in details.

\begin{enumerate}
\item \textit{Deposition}. CdTe adatoms are deposited at a constant rate $R_1$. In this process, a site of substrate is chosen at random and its height increased by a lattice unity.

\item \textit{Diffusion}. The diffusion of CdTe adatoms has an activation energy $E_d = E_{0d} + E_{1d}n$, where $n$ is the number of CdTe-CdTe bounds of the particle. Here, $E_{0d}$ is interpreted as the effective energy for the activation of diffusion and $E_{0d}$ is interpreted as a difference between diffusion activation energies of Si-CdTe and CdTe-CdTe, respectively. To include these interpretations in the model, we adopted the following rules. The CdTe diffusion is tried (but not necessarily implemented) at a rate $R_2$ given by an Arrhenius-like expression [2]
\begin{equation}
R_2= \frac{d^*k_B T}{h}\exp(-E_{0d}/k_BT), 
\end{equation}
where $d^*$ is substrate dimension. However, the diffusion can be implemented or not with probabilities dependent on the local neighborhood of the particle. For each tentative, a site $j$ and one of its nearest-neighbors sites $j^\prime$ are chosen at random. If none CdTe was deposited on the site $j$, the rule is not implemented. Otherwise, the particle hops from $j$ to $j^\prime$ with probability
\begin{equation}
P_{hop}=\left\{\begin{array}{l}
		p^n \mbox{ if } \Delta h \le 0,\\
		p^{n+\Delta h} \mbox{ if } \Delta h > 0
	     \end{array} \right.
\end{equation}
where $\Delta h \equiv h_{j^\prime} -h_j+1$ is the difference between the initial and final height of the particle and $p$ is given by
\begin{equation}
p=\exp\left(-\frac{E_{1d}}{k_B T}\right).
\end{equation}
The upward diffusion is allowed but a Ehrlich-Schwoebel barrier for
$\Delta h > 0$ is included without additional parameters. The downward Ehrlich-Schwoebel barriers were not considered in order to include an incorporation mechanism at lower step edges. One expects that the absence of such a mechanism should lead to a growth-instability [24] not observed in the experimental system studied in this work. It is important to note that except by the Ehrlich-Schwoebel barriers, the long time regime of the diffusion model, when the Si substrate is completely covered by CdTe, is equivalent to the classical model for MBE proposed by Das Sarma and Tomborenea [13].

\item \textit{Evaporation}. The evaporation is believed to play an essential role in the growth of CdTe quantum dots. In the present model, the evaporation, with an activation energy $E_v = E_{0v} + E_{1v}n$, consists of the exclusion of the highest deposited CdTe adatom in a randomly chosen column. The physical interpretations of the parameters $E_{0v}$ e $E_{1v}n$ are analogue to those given for the activation energies of the diffusion rules. Similarly to the diffusion, the evaporation occurs at a rate $R_3$ given by a Arrhenius-like expression
\begin{equation}
R_3= \frac{d^*k_B T}{h}\exp(-E_{0v}/k_BT).
\end{equation}
 This process can be implemented or not with probability $q^n$, where
\begin{equation}
q=\exp\left(-\frac{E_{1v}}{k_B T}\right).
\end{equation}
\end{enumerate}
In figure 3, the previously described processes are illustrated.

\begin{figure}[hbt]
\begin{center}
\includegraphics[clip=true,width=8.0cm]{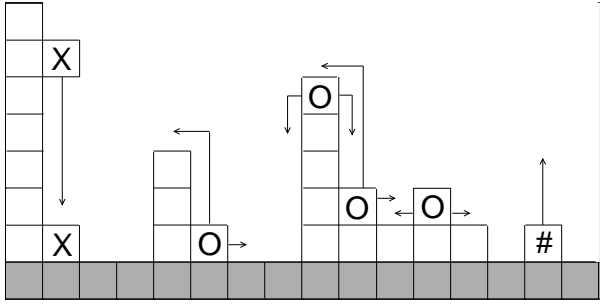}
\end{center}
\vspace{-0.5cm}
\caption{Schematic representation of the processes considered in the model: deposition (X), diffusion (O), and evaporation (\#). The arrows indicate possible particle destinations after the rule implementation. In the evaporation the particle is detached from the substrate. The Si substrate is depicted in gray.}
\label{fig:fig1}
\end{figure}

\textit{Algorithm implementation}. The simulation procedure is the following. At each time step, one of the three processes deposition, diffusion, or evaporation is selected and, if the case, implemented with probabilities $P_1$, $P_2$, and $P_3$, respectively, and the time is incremented by $\Delta t = 1/L$. These probabilities are normalized and proportional to the correspondent rate, i. e., are given by
\begin{equation}
P_i=\frac{R_i}{\sum_{j=1}^{3}R_j}.
\end{equation}

These diffusion and evaporation models include some important physical features. The
hopping probabilities for particles on CdTe is smaller than those for particles on Si, since there are always one out-plane CdTe-CdTe bond for particles on CdTe and, consequently, $n\ge 1$. But, for particles on Si this out-plane component does not exist and $n\ge 0$. This is a desirable feature since the binding energies between CdTe adatoms are more relevant than the ones between CdTe and Si atoms, assuming the VW growth mode. A second feature is the hopping probability from lower to higher columns justified by the difference between the substrate (200 - 300 $^\circ$C) and the vapor source (420 $^\circ$C) temperatures, which is not sufficiently large to neglect this effect [10]. Also, large differences between Si substrate and CdTe layer, namely lattice mismatch, thermal expansion coefficient, crystal structure, and polarity (Si is nonpolar whereas CdTe is polar) [10] also contribute to the jumps from lower Si substrate to higher CdTe layers. Finally, a mechanism included in the diffusion rule is the detachment of CdTe particles from nucleated islands.

\section{Results and Discussions}

\label{results}

Some model parameters can be obtained from the data reported in [10]. In all simulations,
we considered temperatures corresponding to the experiments, namely, 200-400 $^\circ$C and a
constant deposition rate $R_1 = 0.32$ ML/min (see reference [10] for details). The parameter
concerning the activation energies are not experimentally known. Thus, in order to estimate
the order of magnitude of these parameters, we used the bound energies for the CdTe
bulk determined by Oh and Grein [25] through theoretical approaches: $E_{Cd-Te} = 1.03$ eV,
$E_{Te-Te} = 1.06$ eV, and $E_{Cd-Cd} = 0.58$ eV. For the remaining bound energies we used the same values adopted by Pyziak \textit{et al.} [23]: $E_{Cd-Si} = E_{Te-Si} = 0.8$ eV. The the typical activation energy for evaporation of isolated CdTe on Si adatoms is estimated as $E_{0v}\simeq E_{Cd-Si} + E_{Te-Si} = 1.6$ eV, while the typical energy per bound of CdTe-CdTe components is estimated as $E_{1v} \simeq (E_{Cd-Te} + E_{Te-Te} + E_{Cd-Cd})/3 - E_{0v}/2 \simeq 0.1$ eV. Precise informations about the activation energies for the diffusion are not available, but we expect $E_{0d} < E_{0v}$. It is worth
to reinforce that these values are only an estimate of the range in which the parameters
must be varied.

The evolution rules determine two distinct growth conditions: a first one, in which the
Si substrate is not completely covered, and a second one, in which the substrate is entirely
covered. The QDs of CdTe on Si are experimentally observed only in samples with few
monolayers of evaporated material (figure 1). In this regime, it is possible to simulate large
systems with a rigorous statistics in a reasonable computational time. In all simulations
concerning the short time analysis, linear chains containing $L = 10^4$ sites were used and the averages done over $10^3$ independent samples. 

In figure 4, we show profiles generated for distinct amounts of evaporated material, which can be used as a measure of time, at $T = 250~^\circ$C. For short times (figure 4(a)), one can see nucleated CdTe islands containing more than one monolayer with typical height of about 1.5 ML even when less then one monolayer of CdTe was deposited. This mean height value is coherent with the experiments with CdTe grown on Si by HWE  [10] and also consistent with PbTe QDs grown on BaF2(111) by MBE [26]. As additional material is deposited the islands
coalesce generating a rough surface.

\begin{figure}[hbt]
\begin{center}
\includegraphics[clip=true,width=9cm]{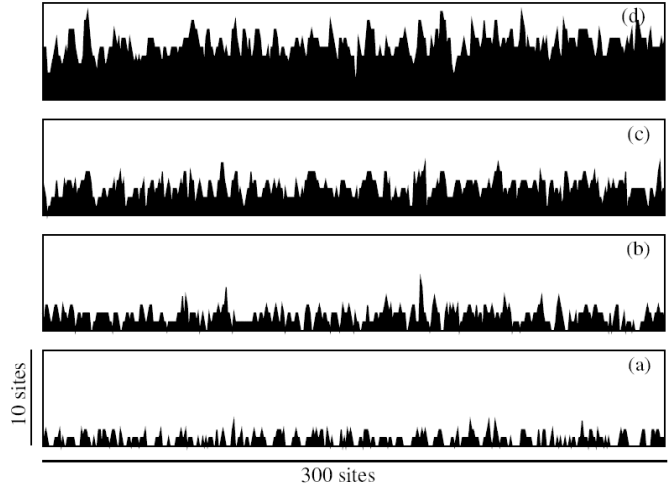}
\end{center}
\vspace{-0.5cm}
\caption{Surface profiles for (a) 0.8 ML, (b) 1.6 ML, (c) 3.2 ML, and (d) 6.4 ML of evaporated material. The parameters used in these simulations were $T = 250~^\circ$C, $E_{1v} = E_{1d} = 0.09$ eV, $E_{0v} = 1.60$ eV, and $E_{0d} = 1.40$ eV.}
\label{pfT250}
\end{figure}

The effect of the temperature is illustrated in figure 5. For the highest temperature ($T =
380~^\circ$C) the desorption becomes dominant, in quantitative agreement with the experiments. For $T \ge 250~^\circ$C, the profiles exhibit QDs that apparently increase in number and decrease in width as the temperature increases. This is a typical feature of the VW growth mode observed in CdTe/Si (figure 2) and similar systems [26]. The main model ingredient that originates the QDs is the difference between diffusion and evaporation rules for CdTe on the initial Si substrate and CdTe on the other previously deposited CdTe. However, at lower temperatures the model suggests a SK (2D) growth mode as indicated by Figs. 5(a) and (b). This hypothetical transition will be careful discussed in the subsequent paragraphs.

\begin{figure}[hbt]
\begin{center}
\includegraphics[clip=true,width=9cm]{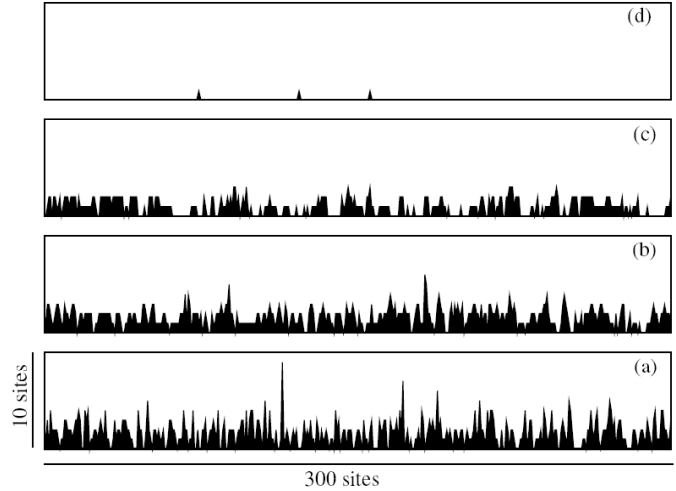}
\end{center}
\vspace{-0.5cm}
\caption{Effects of temperature on the interfaces with 1.6 ML of evaporated CdTe. The temperatures used were (a) $T = 200~^\circ$C, (b) $T = 250~^\circ$C, (c) $T = 300~^\circ$C, and (d) $T = 380~^\circ$C, while the other parameters were $E_{1v} = E_{1d} = 0.09$ eV, $E_{0v} = 1.60$ eV, and $E_{0d} = 1.40$ eV.}
\label{pfT16ML}
\end{figure}

The height and width distributions of the QDs were evaluated for distinct temperatures.
In order to determine the QD heights and widths each dot was fitted by a quadratic polynomial
null at both of its extremities. The heights are assumed as the maximum of the
polynomial. A QD and the corresponding parabolic fits are shown in figure 6. This procedure
intends to produce a more trustworthy variety of dots sizes.

\begin{figure}[hbt]
\begin{center}
\includegraphics[clip=true,width=8cm]{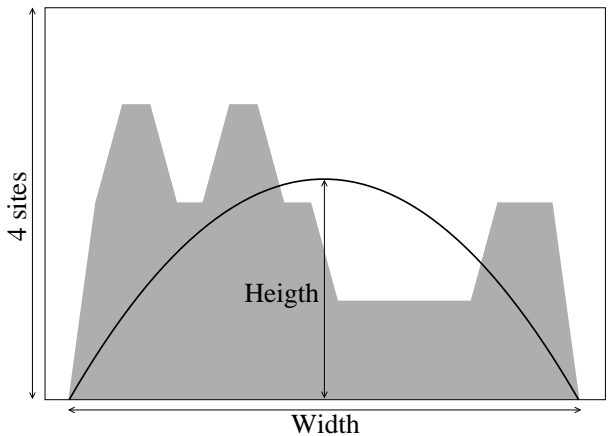}
\end{center}
\vspace{-0.5cm}
\caption{Determination of the QD height and width. Both, the QD (gray) and the quadratic fit (line) are shown.}
\label{quadratic}
\end{figure}

In Fig 7, the distributions obtained for distinct temperatures are drawn. On one hand,
the height distributions were fitted by two-peak Gaussian curves. Such bimodal distributions
are commonly observed in many systems undergoing both SK [5, 27,29] and VW [10, 26]
growth modes. As temperature increases, the curves become sharper and the corresponding
peaks shift to smaller heights. In other words, the number of small QDs increases while the
number of the larger ones decreases. In particular, this qualitative behavior was observed
for CdTe/Si system. On other hand, the QD width distributions have a similar qualitative
behavior (figure 7(b)), but the data are poorly fitted by two-peak Gaussian curves. Indeed, the distributions decay exponentially for large width values, as can be observed in the inset of figure 7(b). The same asymptotic decay was found for the height distributions.

\begin{figure}[hbt]
\begin{center}
\includegraphics[clip=true,width=8.0cm]{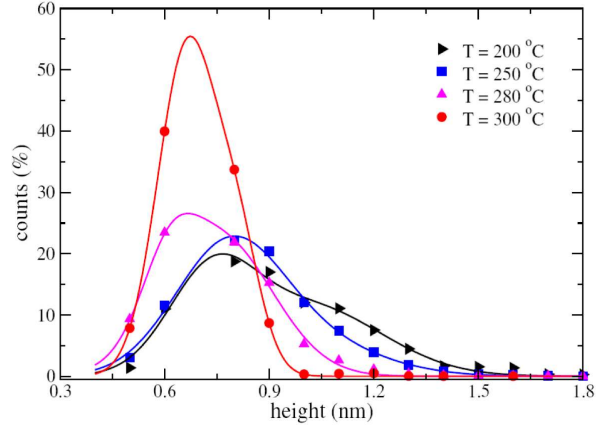}\\(a)
\end{center}
\begin{center}
\includegraphics[clip=true,width=8.0cm]{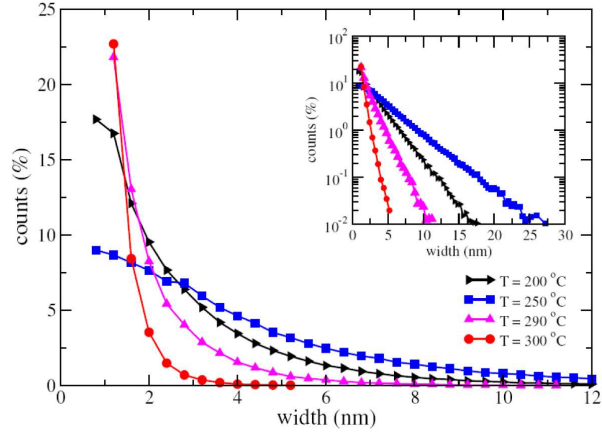}\\(b)
\end{center}
\vspace{-0.5cm}
\caption{(a) Height and (b) width distributions for distinct temperatures. In inset, width distributions are drawn in semi-log plots becoming neat the asymptotically exponential decays. The solid lines correspond to the nonlinear fits. 1.6 ML of evaporated material, $E_{0v} = 1.65$ eV, $E_{0d} = 1.3$ eV, and $E_{1v} = E_{1d} = 0.09$ eV were used in all curves.}
\label{histo}
\end{figure}

An important aspect observed in the width distributions is the curve inversion as the
temperature increases (figure 7(b)). Comparing the distributions for $T = 200~^\circ$C and $T = 250~^\circ$C, the first curve is sharper than the second one. But, comparing the distributions for $T = 250~^\circ$C and  $T = 300~^\circ$C, for example, the opposite is observed. We associated this behavior to a transition from SK to VW growth mode. In figure 8, the mean QD width and height are plotted as functions of temperature for three distinct $E_{0v}$ values. The QD width increases in the low temperature regime, a signature of the SK growth mode, and decreases for the high ones, a feature of the VW growth mode. In CdTe/Si experiments only the decreasing regime of the QD mean width was observed. However, the mean height is always a decreasing function of temperature in agreement with VW mode. In figure 8(c) the mean QD width as a function of temperature is shown for several activation energies for diffusion. This figure evidences the influence of the $E_{0d}$ parameter on the range in which VW-like growth regime is observed. Notice that these data partially disagree with the experiments because the growing mean width regime was not experimentally observed in the interval $200 - 300~^\circ$C (figure 2). However, as $E_{0d}$ decreases the temperature range corresponding to the VW-like growth mode increases, leading to a better agreement with the experiments. Additional experiments at lower temperatures should be executed in order to verify if SK-like growth is also present in CdTe/Si systems. Also, simulations using 2D substrates must be done in order to improve the accuracy of simulations.

\begin{figure}[hbt]
\begin{center}
\includegraphics[clip=true,width=11cm]{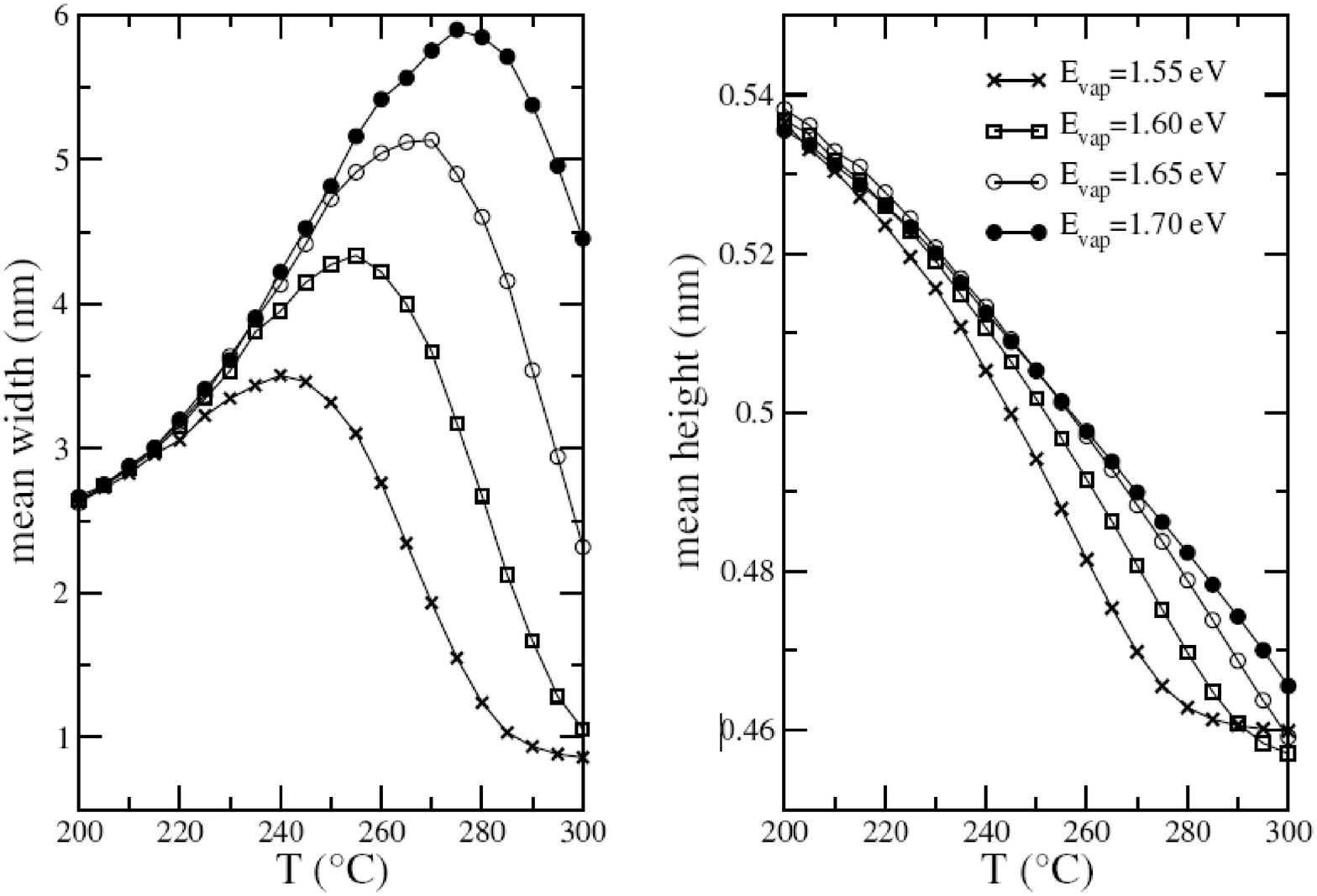}\\
(a)~~~~~~~~~~~~~~~~~~~~~~~~~~~~~~~~~~~~~~~~~~~(b)\\
\includegraphics[clip=true,width=9cm]{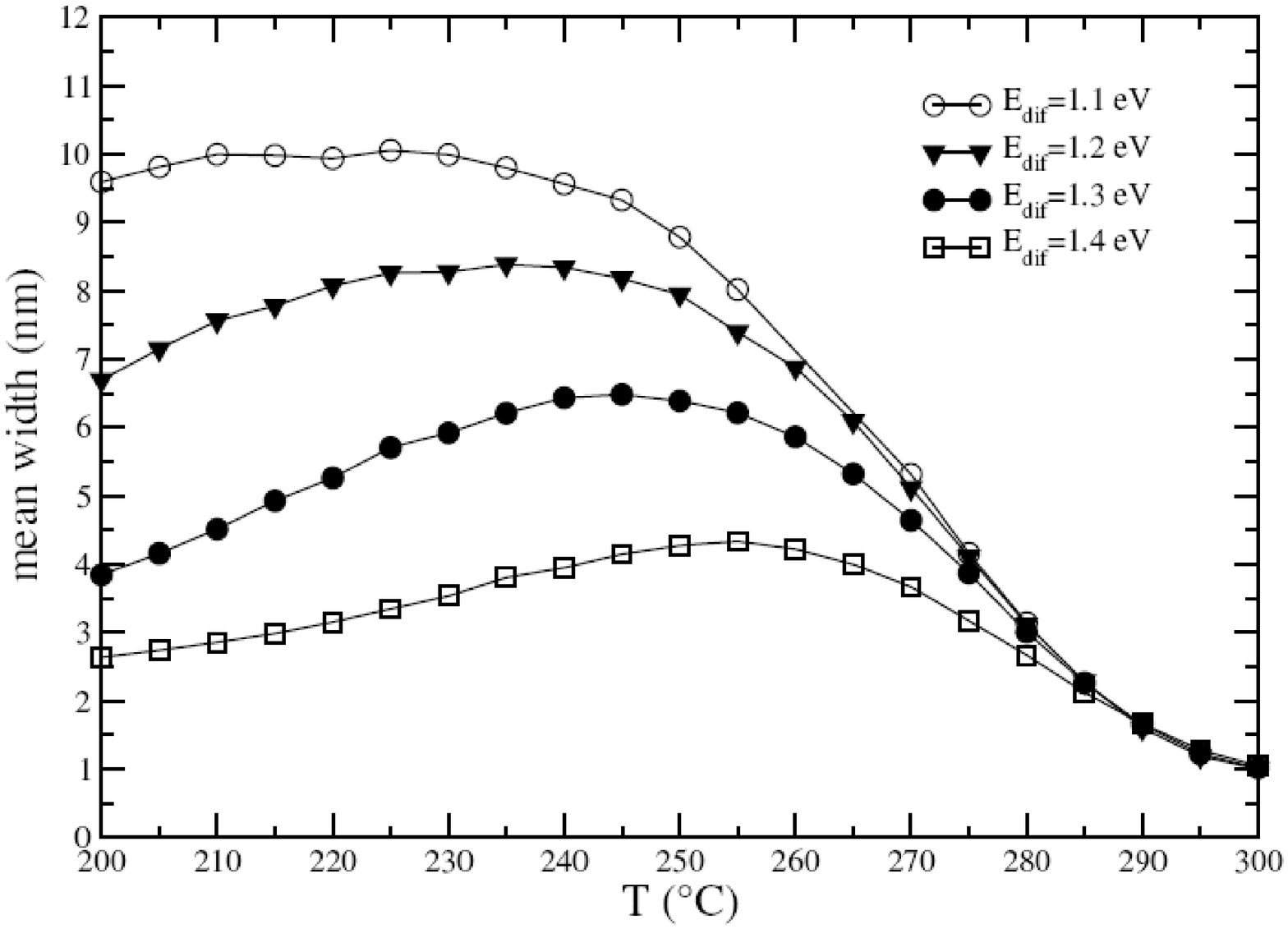}\\
(c)
\end{center}
\vspace{-0.5cm}
\caption{ Mean dot (a) width and (b) height as functions of temperature for distinct activation energies for evaporation, and (c) the width distribution for distinct activation energies for diffusion. 1.6 ML of evaporated material and $E_{1v} = E_{1d} = 0.09$ eV were used in all simulations. Also, $E_{0d} = 1.3$ eV was used in (a) and (b) while $E_{0v} = 1.6$ eV was used in (c).}
\label{skvw}
\end{figure}

\section{Conclusions}
\label{conclusions}

In the present work we studied a one dimensional model for the growth of CdTe quantum
dots (QDs) on Si substrates. The experimental system follows the Volmer-Weber (VW)
growth mode, with nucleation of CdTe dots on the Si surface [10] for less than one monolayer
of deposited material. AFM analysis of the samples showed that the size and density of QDs
can be controlled by properly adjusting the growth temperature. These quantities exhibit
opposite behavior when compared to that observed in systems following Stranski-
Krastanow (SK) growth mode used in almost the totality of the growth of semiconductor
nanostructures. In the present model, CdTe adatoms are deposited on a one-dimensional
Si substrate. The model includes thermally activated diffusion e evaporation of
CdTe atoms. The central feature of the model are the differences among the diffusion and
evaporation rules for CdTe particles on Si substrate and CdTe particles on other previously
deposited CdTe layers.

Even this over-simplified one-dimension model revealed several fundamental features
present in the experiments. In the range of temperature corresponding to that of the experiments, the patterns exhibit island nucleation with width and height distributions in
qualitative agreement with those observed in the experiments. For high temperatures, the
distributions shift to smaller sizes and the number of smaller dots increases while the number of larger ones decreases, a feature which agrees with the experimental observations and has been associated to a characteristics of the VW growth mode. However, the model predicts an inversion of this behavior for lower temperatures, a feature characteristic of
the SK growth mode. Such a transition from VW to SK growth mode at low temperatures
has not been observed experimentally. It is important to mention that simulations in $2 + 1$
dimensions must developed in order to improve the quantitative comparisons with the
experimental data.

Finally, we would like to mention that the long time scaling analysis of the model through
the roughness concept was done. As we have claimed in section ``Model and Methods'' we
found that the present model asymptotically exhibits the same scaling behavior of the Das
Sarma e Tamborenea (DT) model [13], including a temperature dependent growth exponent and a range of temperature with exponents $\beta = 3/8$, $\alpha = 3/2$, and $z = 4$, corresponding to the universality class of Mullins-Herring equation [1, 2, 13]
\begin{equation}
\frac{\partial h}{\partial t}=-\nabla^4 h+ \eta(\mathbf{x},t),
\end{equation}
where $\eta(\mathbf{x},t)$ is a uncorrelated noise.
Since the correct universality class of the DT model is an open question and we do not have
any experimental data concerning this long time regime for a comparison, these results
were omitted for sake of brevity.

\begin{acknowledgments}
We thank M. L. Martins and J. A. Redinz for critical reading of the manuscript and J.
G. Moreira and S. G. Alves for fruitful comments. This work was supported by the CNPq,
FAPEMIG, and FINEP Brazilian agencies.

\end{acknowledgments}

\end{document}